\documentclass[superscriptaddress,twocolumn,tightenlines]{revtex4-1}
\usepackage{amsthm}
\usepackage{amsmath}
\usepackage{bbm, dsfont}
\usepackage{graphicx}
\usepackage[usenames,dvipsnames]{color}
\usepackage[colorlinks=true,citecolor=magenta,urlcolor=blue]{hyperref}
\usepackage[table]{xcolor}
\usepackage{colortbl}
\usepackage{color}
\usepackage{multirow}
\usepackage{hhline}
\usepackage{tabu}
\usepackage{epstopdf}
\usepackage{booktabs}
\usepackage[normalem]{ulem}

\usepackage[textwidth=3cm, textsize=footnotesize]{todonotes}
\newcolumntype{C}[1]{ >{ \centering\arraybackslash}p{#1}}

\newcommand{\be}{\begin{equation}}
\newcommand{\ee}{\end{equation}}

\newcommand{\del}[1]{}

\renewcommand\figurename{Fig.}

\newcommand{\sket}[1]{{\ensuremath{\lvert#1\rangle}}}
\newcommand{\lket}[1]{{\ensuremath{\left\lvert#1\right\rangle}}}
\newcommand{\ket}[1]{\if@display\lket{#1}\else\sket{#1}\fi}

\newcommand{\sbra}[1]{{\ensuremath{\langle#1\rvert}}}
\newcommand{\lbra}[1]{{\ensuremath{\left\langle#1\right\rvert}}}
\newcommand{\bra}[1]{\if@display\lbra{#1}\else\sbra{#1}\fi}

\newcommand{\sbraket}[2]{{\ensuremath{\langle#1\rvert#2\rangle}}}
\newcommand{\lbraket}[2]{{\ensuremath{\left\langle#1\!\left\rvert\vphantom{#1}#2\right.\!\right\rangle}}}
\newcommand{\braket}[2]{\if@display\lbraket{#1}{#2}\else\sbraket{#1}{#2}\fi}

\newcommand{\sketbra}[2]{{\ensuremath{\lvert #1\rangle\!\langle #2\rvert}}}
\newcommand{\lketbra}[2]{{\ensuremath{\left\lvert #1\right\rangle\!\!\left\langle #2\right\rvert}}}
\newcommand{\ketbra}[2]{\if@display\lketbra{#1}{#2}\else\sketbra{#1}{#2}\fi}

\definecolor{gray4}{gray}{0.8}
\definecolor{gray2}{gray}{0.6}

\begin{document}
\title{Security proof for a simplified BB84-like QKD protocol}

\author{Davide Rusca}\email{davide.rusca@unige.ch}
\affiliation{Group of Applied Physics, University of Geneva, Chemin de Pinchat 22, CH-1211 Geneva 4, Switzerland}
\author{Alberto Boaron}
\affiliation{Group of Applied Physics, University of Geneva, Chemin de Pinchat 22, CH-1211 Geneva 4, Switzerland}
\author{Marcos Curty}
\affiliation{EI Telecomunicaci\'on, Dept. of Signal Theory and Communications, University of Vigo, E-36310, Spain}
\author{Anthony Martin}
\affiliation{Group of Applied Physics, University of Geneva, Chemin de Pinchat 22, CH-1211 Geneva 4, Switzerland}
\author{Hugo Zbinden}
\affiliation{Group of Applied Physics, University of Geneva, Chemin de Pinchat 22, CH-1211 Geneva 4, Switzerland}

%%%%%%%%%%%%%%%%%%%%%%%%%%%%%%%%%%%%%%%%%%%%%%%%%%%%%%%%%%%%%%%%%%%%%%%%%%%%%%%%
\begin{abstract}
The security of quantum key distribution (QKD) has been proven for different protocols, in particular for the BB84 protocol. It has been shown that this scheme is robust against eventual imperfections in the state preparation, and sending only three different states delivers the same secret key rate achievable with four states. In this work, we prove, in a finite-key scenario, that the security of this protocol can be maintained even with less measurement operators on the receiver.  This allows us to implement a time-bin encoding scheme with a minimum amount of resources.
\end{abstract}
%%%%%%%%%%%%%%%%%%%%%%%%%%%%%%%%%%%%%%%%%%%%%%%%%%%%%%%%%%%%%%%%%%%%%%%%%%%%%%%%

\maketitle

%%%%%%%%%%%%%%%%%%%%%%%%%%%%%%%%%%%%%%%%%%%%%%%%%%%%%%%%%%%%%%%%%%%%%%%%%%%%%%%%
\section{Introduction}
%%%%%%%%%%%%%%%%%%%%%%%%%%%%%%%%%%%%%%%%%%%%%%%%%%%%%%%%%%%%%%%%%%%%%%%%%%%%%%%%

%Main 
The most popular protocol used in quantum key distribution (QKD) is without any doubt the BB84 protocol, firstly presented by Bennett and Brassard in 1984 \citep{Bennett1984}. The security of this protocol against general attacks has been proven in various scenarios \cite{shor2000,lo1999,mayers2001,koashi2009,koashi2003}. A more realistic scenario of imperfect sources (state preparation errors) was considered at first by Gottesman, Lo, L\"{u}tkenhaus and Preskill (GLLP) \cite{gottesman2004}. Afterwards Tamaki et al. \cite{tamaki2014} proved that the security could be achieved also in a loss tolerant scenario (Eve cannot use the loss of the channel to enhance her attack in the presence of state preparation imperfections). This work also demonstrated that not all four BB84 states are actually needed and an equal secret key rate (SKR) can be achieved with only three prepared states.

However in all these security proofs of the BB84 protocol, it was always considered the possibility of measuring, at Bob's side, the received states in the Z and X basis. 
In our paper we relax this condition showing, in a finite-key scenario, that projections on only three states at Bob's side are enough to precisely estimate all security parameters. This protocol simplification is then applied to a recent time-bin encoding scheme that allows for a simple and practical experimental implementation. For instance our protocol can be implemented by using only one modulator at Alice's side and two detectors at Bob's side~\cite{boaron2017}.
Although it is true that even in a standard BB84 protocol, any experimental scheme with only two detectors could be implemented thanks to active detection basis choice or detection multiplexing, this simplification always comes with additional practical limitations. 
For instance the active choice of basis requires an active modulator which increases the complexity of the system and introduces additional insertion loss. On the other hand, temporal multiplexing the output increases the complexity, the loss and possibly reduces the maximum achievable repetition rate. Our scheme, however, is implementable with passive basis choice and can exploit the maximum rate of acquisition of the system \cite{boaron2018}.   

%%%%%%%%%%%%%%%%%%%%%%%%%%%%%%%%%%%%%%%%%%%%%%%%%%%%%%%%%%%%%%%%%%%%%%%%%%%%%%%%%%
\section{Three-state BB84 and simplified measurement at Bob's side}
%%%%%%%%%%%%%%%%%%%%%%%%%%%%%%%%%%%%%%%%%%%%%%%%%%%%%%%%%%%%%%%%%%%%%%%%%%%%%%%%%%

In our scheme we suppose that Alice sends to Bob two states in the basis Z and only one in the basis X, as in the three-state BB84 protocol \cite{molotkov1996,molotkov1998,shi2000,fung2006}. The Z basis (data line) is used to exchange the secret key and the X basis (monitoring line) has the purpose of estimating the information leaked to a third malicious party (Eve). However, we allow Bob to measure the incoming signals in the Z basis or project them onto only one state in the X basis, taking always into account that a possible measurement result is the no-detection event. In this scenario, in the monitoring line, Alice sends one eigenstate ($\ket{+} = \left(\ket{0} + \ket{1}\right)/\sqrt{2}$) of the X basis and Bob measures only the state orthogonal to it ($\ket{-} = \left(\ket{0} - \ket{1}\right)/\sqrt{2}$). 

With these premises, the protocol can be presented as Alice sends to Bob three possible states, i.e. $\ket{0}, \ket{1}$ and $\ket{+}$. Since the channel or the adversary can introduce loss, a no detection event is represented by the state $\ket{\emptyset}$. As in most QKD security analysis, we conservatively assume that Eve can completely control this eventuality. The only thing that limits her is the basis independent detection efficiency condition. This means that Eve is unable to control the efficiency of detection depending on Bob's basis choice. 

In the GLLP security analysis \cite{gottesman2004} the phase error rate ($e_x$) is given by:

\begin{equation}\label{eq_ex1}
e_x = \frac{p(-|+) + p(+|-)}{p(-|+) + p(+|+) + p(+|-) + p(-|-)},
\end{equation}
where $p(j_B|j_A)$ is the probability of Bob detecting the state $j_B$ when Alice sent the state $j_A$.
In our protocol we have no  direct way to measure the probabilities where either Alice sends the state $\ket{-}$ or where Bob measures the state $\ket{+}$. However in a scenario where an attacker is limited to collective attacks and the probabilities of choosing the bases Z and X are (for the moment) equal, it can be shown that the phase error ($e_x$) is estimated precisely by the available probabilities:

\begin{multline}\label{eq_ex2}
e_x = \frac{p(-|+)}{\sum\limits_{i,j=0,1} p(i|j)} \\ 
+ \mathcal{M}\left(1 + \frac{p(-|+) - \sum\limits_{i=0,1}{\left( p(-|i) + p(i|+)\right)}}{\sum\limits_{i,j=0,1} p(i|j)}\right),
\end{multline}
where $\mathcal{M}(y) \equiv \max(0,y)$.

To obtain Eq.~(\ref{eq_ex2}), note that in the framework considered the attack of Eve can be modelled with a unitary transformation in the Hilbert space $\mathcal{H}_A\otimes\mathcal{H}_E$ where $\mathcal{H}_A$ is the space of states sent by Alice and received at Bob's side and $\mathcal{H}_E$ is the space of ancilla's states possessed by Eve. The states take the form of:

\begin{multline}\label{eq_st1}
\mathcal{U}_{AE} \ket{j_A}_A\ket{\phi}_E = \\
 \ket{0}_A\ket{\phi_{j_A}^0}_E +  \ket{1}_A\ket{\phi_{j_A}^1}_E + \ket{\emptyset}_A\ket{\phi_{j_A}^{\emptyset}}_E,
\end{multline}
where $j_A \in \lbrace0,1\rbrace$ and $\ket{\phi_{j_A}^{j'_Z}}_E$ are unnormalized states in Eve's hand. Hereafter we will omit the subsystems labels whenever the context allows it. The transformation $\mathcal{U}_{AE} \ket{\pm}\ket{\phi}$ is just a linear combination of the two previous relations given by Eq.~(\ref{eq_st1}). In the framework of colletive attacks, the eavesdropper is constrained to do the same transformation on each pulse, but she can delay her measurement (by storing her states in a quantum memory) until the classical communication between Alice and Bob has been finished. We remark that the attack considered by Eq.~(\ref{eq_ex1}) is not the most general collective attack possible. In fact, Eve, in principle, could send states to Bob with multiple photons. However the high number of parameters to consider makes an analytical result difficult to calculate. The analysis carried out here, even if not completely general, might be proven to be enough for security once a squashing model \cite{beaudry2008,tsurumaru2008,tsurumaru2010,fung2011,gittsovich2014} for the detection scheme is provided.

In order to prove that Eq.~(\ref{eq_ex1}) and Eq.~(\ref{eq_ex2}) are equivalent, it is sufficient to evaluate the conditional probabilities (after the considered collective attack from the eavesdropper) in case of a perfect BB84, which in our protocol are given by the general expression:

\begin{equation}\label{eq_p}
p(j_B|i_A) = \vert\bra{j_B}\mathcal{U}_{AE} \ket{i_A}\ket{\phi}\vert^2.
\end{equation}

\section{Time-bin encoding}

\begin{figure}
\includegraphics[width = \columnwidth]{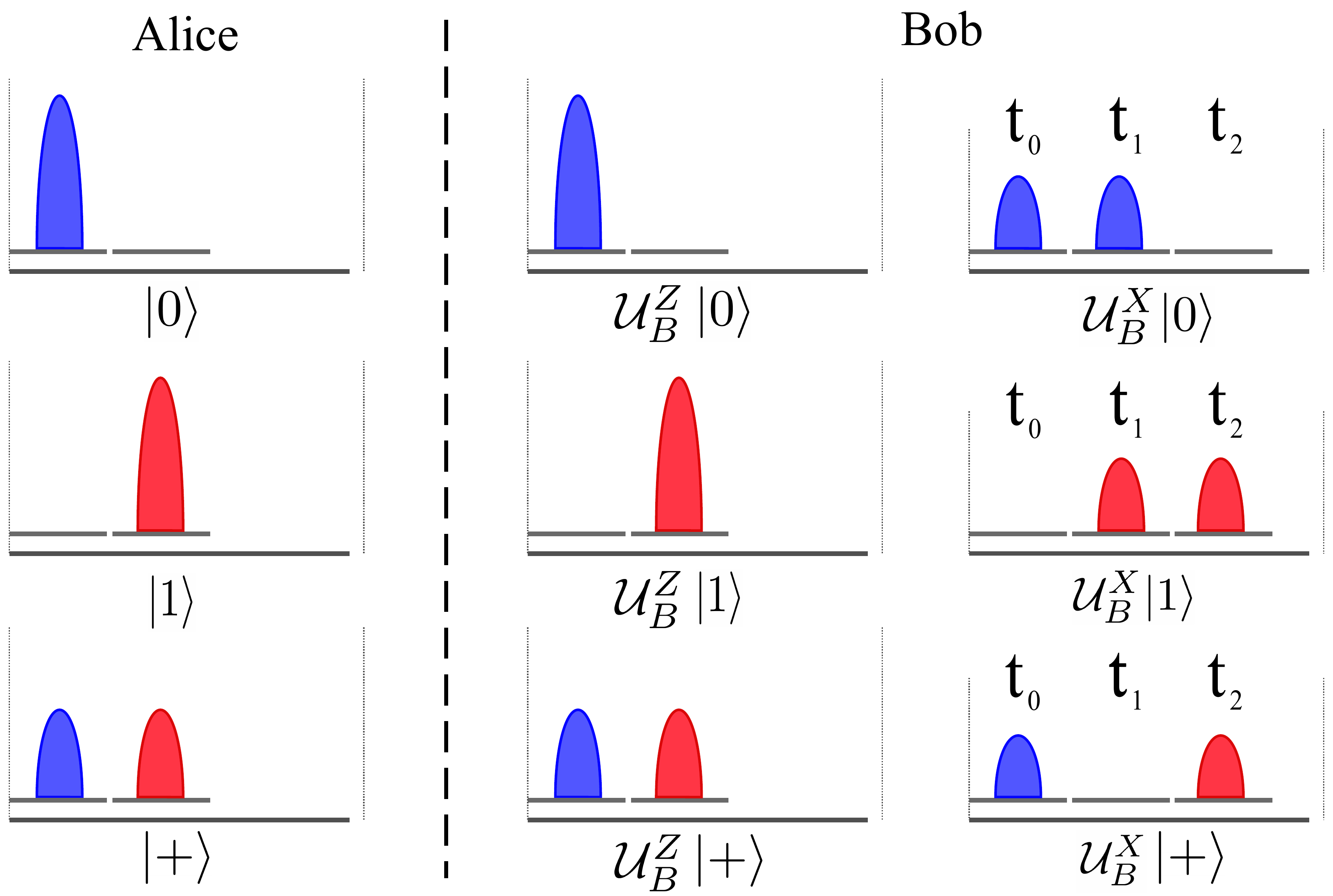}
\caption{\label{fig_States}State generated and measured by Bob in a standard time-bin protocol.}
\end{figure}

The simple implementation presented in the previous section gives already an understanding of the limitation and capabilities of eliminating one of the typical measurement projections from the original BB84 scheme.
Although such analysis is already complete it has some minor drawbacks: the previous formula of the phase error rate (Eq.~(\ref{eq_ex2})) depends on detection events on both the data and monitoring lines. From a practical point of view, it might be more convenient to estimate the phase error rate only using the monitoring basis, where all the bits are usually disclosed between Alice and Bob.

In the detection scheme that we propose, we overcome this limitation. This method is based on a time-bin encoding scheme, where the states sent by Alice corresponds to $\ket{0} = \ket{1}_e\ket{0}_l$, $\ket{1} = \ket{0}_e\ket{1}_l$ and $\ket{+} = \frac{1}{\sqrt{2}}\left(\ket{1}_e\ket{0}_l + \ket{0}_e\ket{1}_l\right)$ (\figurename{~\ref{fig_States}}) where the subscripts $e$ and $l$ represent the early and late time-bins, respectively.

\begin{figure}
\includegraphics[width = \columnwidth]{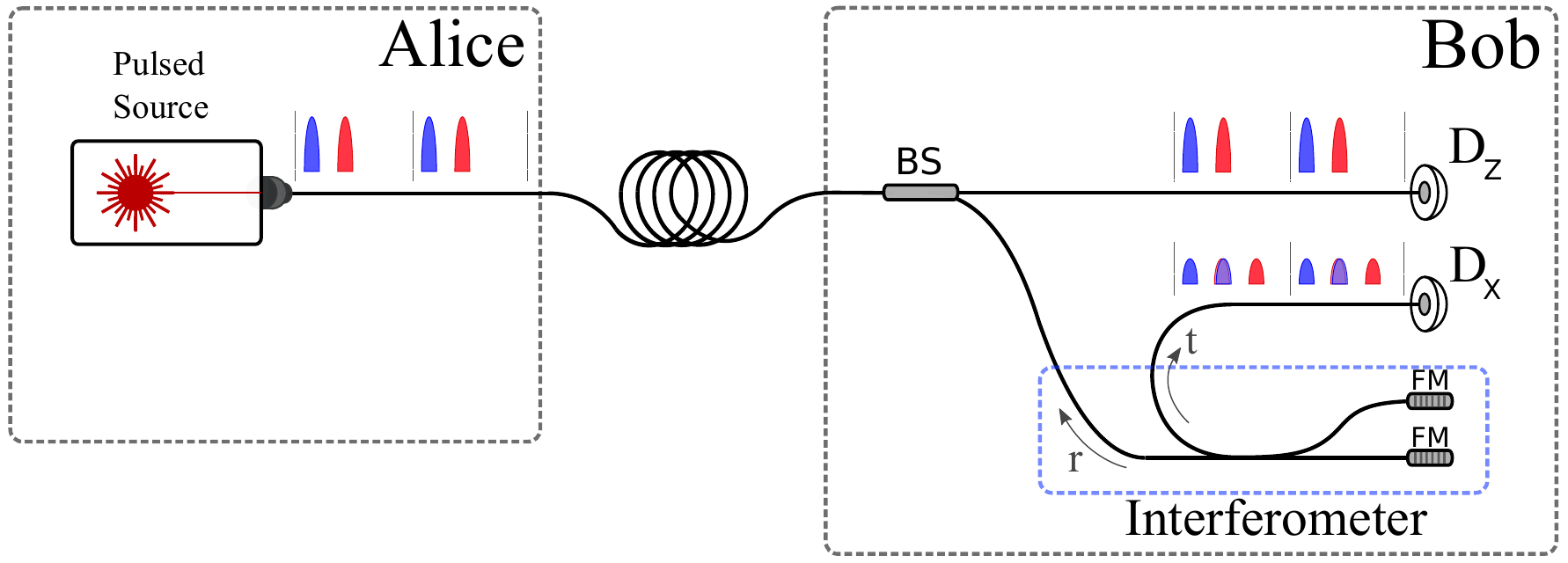}
\caption{\label{fig_Setup}Experimental setup of a three-state BB84 time-bin encoding protocol. Alice has a pulsed laser source and encodes the states in two possible time-bins. Bob, after a passive basis choice done by a beam splitter (BS) can measure in two different bases. In the Z basis, the measurement consists only in measuring the arrival time of the photons. Instead, the X basis consists in measuring the interference of the two time-bins. In this scheme, only one output port of Bob's Michelson interferometer is monitored. The interferometer is composed by a beam splitter and two Faraday mirrors (FM).}
\end{figure}

Bob's detections in the Z basis correspond to a detection in one of the corresponding time-bins, while the monitoring line is composed by an unbalance interferometer as shown in \figurename{~\ref{fig_Setup}}, that allows to measure the coherence between the two time-bins within a state. The action of the optical elements in front of the detectors, in the data and monitoring lines, can be described by the unitary transformations $\mathcal{U}^Z_{B}$ and $\mathcal{U}^X_{B}$ respectively (see \figurename{~\ref{fig_States}}). The data line transformation corresponds, trivially, to the identity, since no optical element appears before the detector.  The monitoring line maps the incoming state into six different possible states. The new states are encoded in two spatial modes, given by the two outputs of the interferometer ($t,r$), and three temporal modes $t_i$ labelled by the subscript $i = 0,1,2$. This detection scheme can be modelled by the unitary transformation $\mathcal{U}^X_{B}$ as following:

\begin{equation}\label{eq_bob}
\begin{array}{l}
\mathcal{U}^X_{B} \ket{0}_A = \frac{1}{2} \left[ \ket{t_0}_B - \ket{t_1}_B\right] + \frac{1}{2} \left[ \ket{r_0}_B + \ket{r_1}_B\right]\\
\mathcal{U}^X_{B} \ket{1}_A = \frac{1}{2} \left[ \ket{t_1}_B - \ket{t_2}_B\right] + \frac{1}{2} \left[ \ket{r_1}_B + \ket{r_2}_B\right]\\
\mathcal{U}^X_{B} \ket{\emptyset}_A = \ket{\emptyset}_B
\end{array}
\end{equation}

Since the only output monitored is the one corresponding to the states $\ket{t_i}$, a projection in one of the $\ket{r_i}$ states results in a no-detection event. Considering Eve's transformation, given by Eq.~(\ref{eq_ex1}), the states received at Bob's monitoring line have the form:

\begin{multline}\label{st_bob}
\mathcal{U}^X_{B}\mathcal{U}_{AE} \ket{j_A}_A\ket{\phi}_E = \\
 \frac{1}{2} \left( \ket{t_0}_B - \ket{t_1}_B + \ket{r_0}_B + \ket{r_1}_B\right)\ket{\phi_{j_A}^0}_E  \\
 + \frac{1}{2}  \left( \ket{t_1}_B - \ket{t_2}_B + \ket{r_1}_B + \ket{r_2}_B\right)\ket{\phi_{j_A}^1}_E \\
 + \ket{\emptyset}\ket{\phi_{j_A}^{\emptyset}}.
\end{multline}

In this scenario the conditional probabilities $p(j_B|i_A)$ can be calculated as in Eq.~(\ref{eq_p}) and the phase error rate can be expressed as:

\begin{multline}\label{eq_ex3}
e_x = \frac{p(t_1|+)}{2\sum\limits_{i=0,1}\sum\limits_{j=0,2} p(t_j|i)}\\
 + \mathcal{M}\left(1 + \frac{\frac{1}{2} (p(t_1|+) - p(t_1|Z)) -  p(t_{side}|+)}{\sum\limits_{i=0,1}\sum\limits_{j=0,2} p(t_j|i)} \right),
\end{multline}
where $p(t_1|Z) = p(t_1|0) + p(t_1|1)$ and $p(t_{side}|+) = p(t_0|+) + p(t_2|+)$ (see Sec.~\ref{Security} for the derivation of Eq.~(\ref{eq_ex3})).

It can be easily verified that all the terms in Eq.~(\ref{eq_ex3}) depend only on the possible detections in the monitoring line at Bob's side, $b \in \lbrace t_0,t_1,t_2\rbrace$. The reason why this is possible becomes clear when we explicit the POVM elements of the monitoring line measurement, $M_b$:

\begin{equation}\label{eq_POVM}
\begin{array}{l}
M_{t_0} = \frac{1}{4}\ket{0}\bra{0} = \frac{1}{4}\ket{1,0}\bra{1,0},\\
M_{t_1} = \frac{1}{2}\ket{-}\bra{-} = \frac{1}{4}(\ket{1,0}-\ket{0,1})(\bra{1,0}-\bra{0,1}),\\
M_{t_2} = \frac{1}{4}\ket{1}\bra{1} = \frac{1}{4}\ket{0,1}\bra{0,1},\\
M_{\emptyset} = \frac{1}{2}\mathbbm{1} + \frac{1}{2}\ket{0,0}\bra{0,0} + \frac{1}{4}\left(\ket{0,1}\bra{1,0} + \ket{1,0}\bra{0,1}\right).
\end{array}
\end{equation}
From this expression we can directly see that the POVM's elements $M_{t_0}$ and $M_{t_2}$ correspond to projections on the Z basis, with the exception of a renormalization factor. This opens the possibility of a further simplification of our scheme. Indeed, what we defined in our scheme as the monitoring line, could be used both as the monitoring and data lines. The side peaks of the interferometer (i.e. the pulses at the times $t_0$ and $t_2$) could be regarded directly as the data line and the estimation of the phase error rate would be unchanged. This gives the possibility of using only one detector for the whole scheme.

\section{Efficient encoding scheme}

\begin{figure}
\includegraphics[width = \columnwidth]{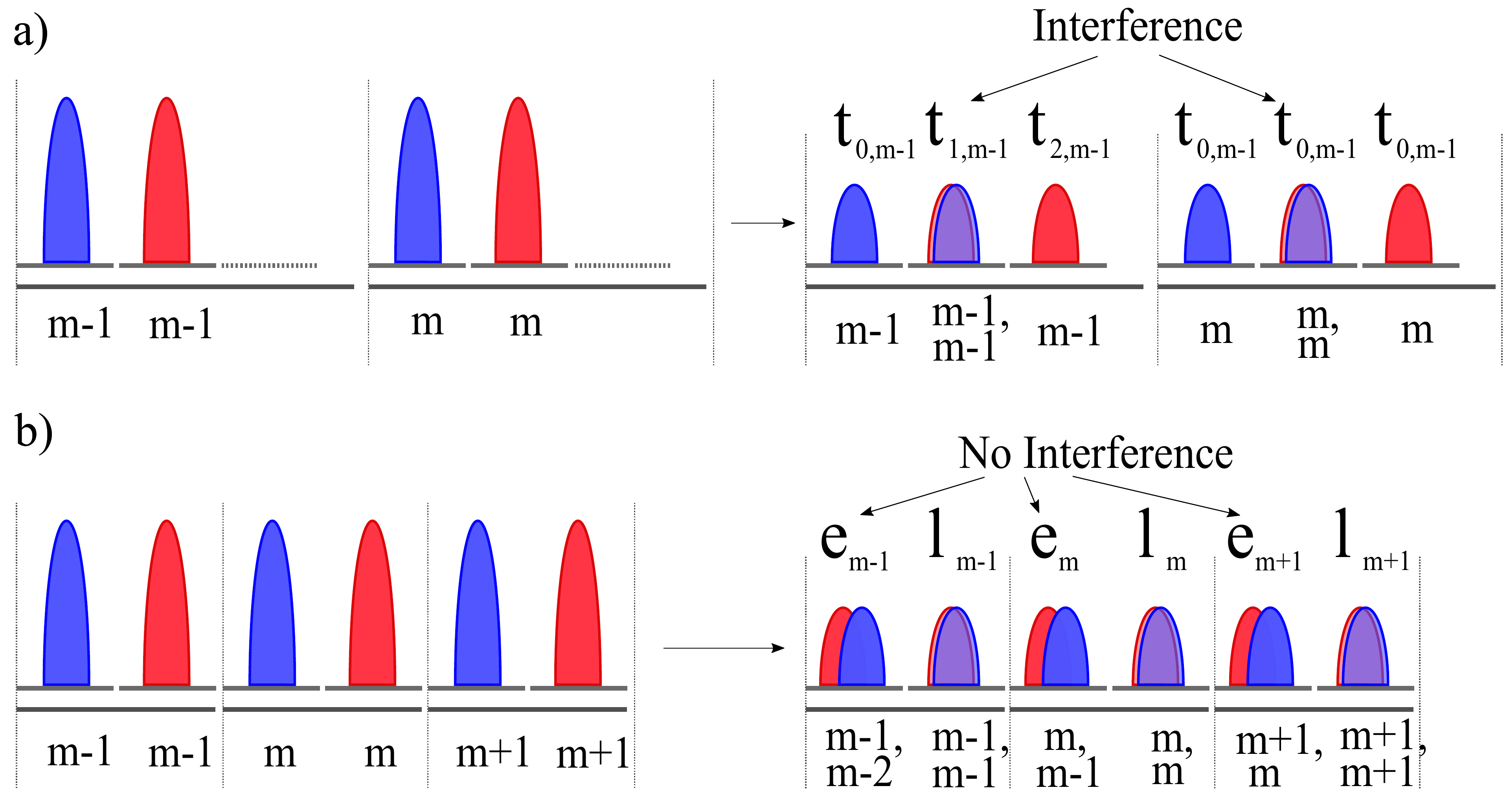}
\caption{\label{fig_time}\textbf{a)} The simplest implementation of a time-bin encoding scheme. The time difference between two different states is greater than three times the time-bin duration. In this way the three possible detection times on the monitoring line are not overlapped. \textbf{b)} The most efficient time bin scheme: in this case the time difference between two different states is exactly the total time duration of a state (two time-bins). In this case the early detection time-bin in the monitoring line depends on both the current pulse and the previous one.}
\end{figure}

In a time-bin encoding scheme one important parameter to consider is the size of the temporal mode. Once this parameter is fixed, the minimum temporal width of the qubit state is limited at two temporal modes (see \figurename{~\ref{fig_time}}). Although this solution (Fig.~\ref{fig_time}b)) is the most efficient in terms of qubit repetition rate it raises some difficulties in our analysis. The problem appears in the monitoring line where, if we define a pair of states sent by Alice as $\ket{j_{m-1}}\ket{j'_m}$, we have that the detections $\ket{t_{2,m-1}}$ and $\ket{t_{0,m}}$ are indistinguishable (they overlap in time). This fact does not allow us to directly measure all the probabilities of the form $p(b|j)$ where $b \in \lbrace t_0,t_2 \rbrace$ and $j \in \lbrace0,1,+\rbrace$.
However, assuming that two identically encoded states are indistinguishable one from the other, we can evaluate the sum of the detection probabilities corresponding to the states $\ket{t_0}$ and $\ket{t_2}$ when two consecutive identical states are prepared.
More precisely, we define in the monitoring line the possible detections as early ($\ket{e_m}$) and late ($\ket{l_m}$) where this two time-bins have the same temporal separation as the preparation state (see Fig.~\ref{fig_time}b)). It can be easily verified that the late detection $\ket{l_m}$ depends only on the state sent in the $m^{th}$ round. On the other hand the detection in the early bin $\ket{e_m}$ depends both on the current $m^{th}$ round and on the previous one $(m-1)^{th}$. When we send two identical states $\ket{j}$ one after the other, i.e. $\ket{j_{m-1}}\ket{j_m} \equiv \ket{j}\ket{j}$, we have the following relation:

\begin{equation}\label{eq_pjj}
p(t_0|j) + p(t_2|j) = p(e|j,j),
\end{equation}
where on the left side the probabilities $p(t_0|j)$ and $p(t_2|j)$ are the probabilities of detection due only to one preparation state. In this way we can express the phase error rate, given by Eq.~(\ref{eq_ex3}) as follows:

\begin{multline}
e_x = \frac{p(l|+)}{2\sum\limits_{i=0,1} p(e|i,i)} + \mathcal{M}\left(1 + \frac{\frac{1}{2} p(l|+)}{\sum\limits_{i=0,1} p(e|i,i)} \right. \\
\left. - \frac{\frac{1}{2} p(l|Z) + p(e|+,+)}{\sum\limits_{i=0,1} p(e|i,i)}\right),
\end{multline}
where we consider only detection events in the early or late time-bins of the monitoring line and where $p(l|Z) = p(l|0) + p(l|1)$.

The difference between evaluating each probability independently and coupling the states in pairs appears only when we pass from conditional probabilities to joint probabilities  (this passage is convenient for the decoy-state analysis presented in the next section). In fact, to evaluate the joint probabilities we apply the following relations for the early ($e$) and late ($l$) time-bins respectively:

\begin{multline}\label{eq_pjE}
p(e,j,j) = p(e|j,j)p(j,j) = p(e|j,j)p(j)^2,
\end{multline}
and
\begin{equation}\label{eq_pjL}
p(l,j) = p(l|j)p(j).
\end{equation}

We can now rewrite the phase error formula with these new definitions:

\begin{multline}\label{eq_ex4}
e_x = \frac{\alpha}{2} \frac{p(l,+)}{\sum\limits_{i=0,1} p(e,i,i)} + \mathcal{M}\left(1 + \frac{\alpha}{2} \frac{ p(l,+)}{\sum\limits_{i=0,1} p(e,i,i)} \right.\\
\left. - \frac{ \beta p(l,Z) + \alpha p(e,+,+)}{\sum\limits_{i=0,1} p(e,i,i)} \right),
\end{multline}
where $\alpha = p_z^2/(4(1-p_z))$, $\beta = p_z/4$ and $p_z$, $p_x$ are the probabilities that Alice emits a state in the Z and X bases, respectively.

Note that instead of using $p(t_0|+) + p(t_2|+) = p(e|++)$ as indicated in Eq.~(\ref{eq_pjj}), it is possible to use an equivalent relation (in case of perfect state preparation), given by:

\begin{equation}
p(t_0|+) + p(t_2|+) = p(e|0+) + p(e|+1).
\end{equation}
This holds since the state $\ket{0}$ should in principle give no contribution in the detection time-bin $\ket{t_2}$ and the state $\ket{1}$ should behave in the same way for the detection time-bin $\ket{t_0}$. This correction is done in order to increase the available number of events to estimate these probabilities since the probability to send a state in the X basis is usually much lower than the probability to send a state in the Z basis in order to maximize the SKR.

\section{Decoy-state parameter estimation}

The analysis carried out until now considers a single-photon source. Unfortunately such a source, that produces single photons deterministically and with high repetition rate, is not yet available. In QKD and related technologies, this kind of sources is typically replaced by weak coherent pulses, which opens possible side channels exploitable by an eavesdropper, due to the presence of multiphoton pulses~\cite{huttner1995,brassard2000}.

Amongst the different possible solutions to solve this issue~\cite{Scarani2004,Stucki2005,koashi2004,tamaki2009}, the most frequently employed and most practical is the decoy-state method~\cite{Hwang2003,Lo2004,wang2005}, where Alice sends to Bob phase-randomized weak coherent pulses with different intensities in order to bound the number of detections at Bob's side due to single-photons ($D_1$). The security of this protocol has been proven in different kinds of configurations with limited number of intensities~\cite{Ma2005,hayashi2014,Lim2014}. For our analysis we choose the implementation with only one decoy~\cite{Ma2005,rusca2018}. With this analysis we can upper bound the phase error rate associated to the single-photon contributions by the following expression:

%\begin{multline}
%\overline{D_1}(e_x) =\frac{p_z^2}{8(1-p_z)} \frac{\overline{D_1}(n(L,+))}{\underline{D_1}(n(E,00) +  n(E,11))} 
%+\max\left(0,  1 + \left( \frac{p_z^2}{8(1-p_z)} \overline{D_1}(n(L,+))
% -  \frac{p_z}{4} \underline{D_1}(n(L,0) + n(L,1))\right.\right.\\
%\left.\left.  - \frac{p_z}{2(1-p_z)}  \underline{D_1}(n(E,0+) + n(E,+1))\right)/\underline{D_1}(n(E,00) +  n(E,11)) \right).
%\end{multline}

\begin{multline}\label{eq_dec}
\overline{D_1}(e_x) =\frac{\alpha}{2} \frac{\overline{D_1}(n(l,+))}{\underline{D_1}(n(e,ZZ))} \\
+\mathcal{M}\left(1 +  \frac{\alpha}{2} \frac{\overline{D_1}(n(l,+))}{\underline{D_1}(n(e,ZZ))} -  \beta \frac{\underline{D_1}(n(l,0) + n(l,1))}{\underline{D_1}(n(e,ZZ))}\right.\\
\left. -  \alpha \frac{\underline{D_1}(n(e,0+) + n(e,+1))}{\underline{D_1}(n(e,ZZ))}\right),
\end{multline}
where $n(e,ZZ) = n(e,00) + n(e,11)$ and $n(b,j)$ is the number of experimentally observed detections at Bob's side when Alice sent a weak coherent pulse encoded in the $\ket{j}$ state.
In the considered one-decoy state protocol with finete-key corrections~\cite{rusca2018}, two different intensities $k \in \lbrace\mu_1,\mu_2\rbrace$ are chosen for each state. In the finite-key regime, the number of detections associated to single-photon events $D_1(n)$, where $n = \sum n_k$ could be any kind of detection at Bob's side, is bounded by the following equations:

\begin{multline}
\underline{D_1}(n) = \frac{\tau_1\mu_1}{\mu_2(\mu_1-\mu_2)}\left(n_{\mu_2}^- - \frac{\mu^2_2}{\mu^2_1}n_{\mu_1}^+ \right.\\
\left. + \frac{(\mu^2_1-\mu^2_2)}{\mu^2_1}\left(\frac{\overline{D_0}(n)}{\tau_0}\right)\right),
\end{multline}

\begin{equation}
\overline{D_1}(n) = \frac{\tau_1}{\mu_1-\mu_2}\left(n_{\mu_1}^+ - n_{\mu_2}^-\right).
\end{equation}
where $\tau_0$ and $\tau_1$ are the total probabilities to send a vacuum state and a single-photon state respectively; $n_k^\pm$ is the finite-key correction, obtained by using the Hoeffding's inequality~\cite{Hoeffding1963}, of the number of detections due to the state of intensity $k \in \left\lbrace \mu_1, \mu_2\right\rbrace $:
\begin{align}
n_k^\pm &:=   \frac{e^{k}}{p_k}\left(n_k \pm \sqrt{\frac{n}{2}\log\frac{1}{\varepsilon}}\right).
\end{align}
where $p_k$ is the probability to send a state of intensity $k$.
The number of vacuum events $\overline{D_0}(n)$ is estimated using the sequence of states $\ket{01}$ and measuring in the late time bin $\ket{l}$ in the monitoring line:

\begin{multline}
\overline{D_0}(n(b,j,j)) = \\
 \frac{p(j,j)}{p(01)}n(e,01) + \delta\left(\frac{p(j,j)}{p(01)}n(e,01),\varepsilon\right),
\end{multline}
where $\delta(n,\varepsilon) = \sqrt{(n\log\varepsilon^{-1})/2}$.

We chose this sequence because, in case of perfect preparation of the state, the only contribution to this event is a vacuum state detection (where Alice sends a vacuum state and Bob has a detection). In case of possible preparation errors in these two states, the detections considered could depend also on non-vacuum events, however this poses no treat to the security of the protocol since the quantity considered is still an upper bound on the considered events.

Then the SKR is given by the following formula~\cite{rusca2018}:

\begin{multline}\label{eq_skr}
l \leq  \underline{D_0}_Z + \underline{D_1}_Z(1-h(\overline{D_1}(e_z))) - \lambda_\text{EC}\\
 - 6\log_2(19/\epsilon_\text{sec}) - \log_2(2/\epsilon_\text{cor}),
\end{multline}
where $\underline{D_0}_Z$ and $\underline{D_1}_Z$ are the lower bounds of the vacuum events and single-photon events when Alice and Bob choose to send and to measure in the Z basis, $\overline{D_1}(e_z)$ is the upper bound on the phase error rate associated to the single-photon contributions (obtained following the procedure in \citep{rusca2018} and we omit it here for simplicity), $\lambda_\text{EC}$ is the number of disclosed bits in the error correction stage and $\epsilon_\text{sec}$ and $\epsilon_\text{cor}$ are the secrecy and correctness parameters. The phase error rate in the Z basis $\overline{D_1}(e_z)$ is obtained from the error rate in the monitoring line $\overline{D_1}(e_x)$ by a finite key correction given by the formula~\cite{Lim2014}:

\begin{multline}
\overline{D_1}(e_z) \leq \overline{D_1}(e_x)\\
+\gamma\left(\varepsilon_\text{sec},\overline{D_1}(e_x),\underline{D_1}_Z,\underline{D_1}(n(e,ZZ))\right),
\end{multline}
where:
\begin{multline}
\gamma\left(a,b,c,d\right)\\
 = \sqrt{\frac{(c+d)(1-b)b}{cd\log2}\log_2\left(\frac{c+d}{cd(1-b)b}\frac{21^2}{a^2}\right)}.
\end{multline}

\section{Security proof} \label{Security}

In this section we show the procedure to obtain Eq.~(\ref{eq_ex3}). According to Eq.~(\ref{eq_st1}), when Alice sends a state in the Z basis we have that:

\begin{equation}
\begin{array}{l}
\mathcal{U}_{AE} \ket{0}\ket{\phi} = \ket{0}\ket{\phi_0^0} +  \ket{1}\ket{\phi_0^1} + \ket{\emptyset}\ket{\phi_0^{\emptyset}},\\
\mathcal{U}_{AE} \ket{1}\ket{\phi} = \ket{0}\ket{\phi_1^0} +  \ket{1}\ket{\phi_1^1} + \ket{\emptyset}\ket{\phi_1^{\emptyset}},
\end{array}
\end{equation} 
and for the $\mathsf{X}$ basis:

\begin{multline}
\mathcal{U}_{AE} \ket{\pm}\ket{\phi} = \frac{1}{\sqrt{2}}\left(\ket{0} ( \ket{\phi_0^0} \pm \ket{\phi_1^0} )\right.  \\
\left. + \ket{1} (\ket{\phi_0^1} \pm \ket{\phi_1^1} ) + \ket{\emptyset} (\ket{\phi_0^{\emptyset}} \pm \ket{\phi_1^{\emptyset}} )\right),
\end{multline}
where $\ket{\phi_i^j}$ for $i,j=0,1,\emptyset$ are Eve's quantum states (not necessarily normalized). By using the state transformation at Bob's monitoring line given by the Eq.~(\ref{eq_bob}), we can calculate the probability that Bob detects $\ket{b}$, with $b\in \{\emptyset,t_0,t_1,t_2\}$, when Alice prepares $\ket{a}$, with $a\in \{0,1,+\}$. In fact the states received by Bob after Eve's attack have the form given by Eq.~(\ref{st_bob}).

Knowing the explicit form of these states and using Eq.~(\ref{eq_p}), we can express all the possible conditional probabilities in the monitoring line (corresponding to measurable events) in terms of Eve's ancilla states:

\begin{equation}\label{eq_pxz}
\begin{array}{l}
p(t_0|0)=  \frac{1}{2} \braket{\phi_0^0}{\phi_0^0},\\
p(t_1|0) = \frac{1}{2} \left[ \braket{\phi_0^0}{\phi_0^0} + \braket{\phi_0^1}{\phi_0^1} - 2 Re( \braket{\phi_0^0}{\phi_0^1})\right],\\
p(t_2|0) = \frac{1}{2} \braket{\phi_0^1}{\phi_0^1},\\
p(\emptyset|0) = \braket{\phi_0^{\emptyset}}{\phi_0^{\emptyset}},\\
p(t_0|1)=  \frac{1}{2} \braket{\phi_1^0}{\phi_1^0},\\
p(t_1|1) = \frac{1}{2} \left[ \braket{\phi_1^0}{\phi_1^0} + \braket{\phi_1^1}{\phi_1^1} - 2 Re( \braket{\phi_1^0}{\phi_1^1})\right],\\
p(t_2|1) = \frac{1}{2} \braket{\phi_1^1}{\phi_1^1},\\
p(\emptyset|1) = \braket{\phi_1^{\emptyset}}{\phi_1^{\emptyset}},
\end{array}
\end{equation}
and 
\begin{equation}\label{eq_pxx}
\begin{array}{ll}
p(t_0|+) &=  \frac{1}{4}\left[ \braket{\phi_0^0}{\phi_0^0} + \braket{\phi_1^0}{\phi_1^0} + 2 Re(\braket{\phi_0^0}{\phi_1^0} ) \right],\\
p(t_1|+) &= \frac{1}{4} \left[\braket{\phi_0^0}{\phi_0^0} + \braket{\phi_1^0}{\phi_1^0} + \braket{\phi_0^1}{\phi_0^1} + \braket{\phi_1^1}{\phi_1^1}\right.\\
& - 2Re(\braket{\phi_0^0}{\phi_0^1}) +2Re(\braket{\phi_0^0}{\phi_1^0})-2Re(\braket{\phi_0^0}{\phi_1^1})\\
&\left.  -2Re(\braket{\phi_0^1}{\phi_1^0}) +2Re(\braket{\phi_0^1}{\phi_1^1}) -2Re(\braket{\phi_1^0}{\phi_1^1}) \right],\\
p(t_2|+) &= \frac{1}{4}\left[ \braket{\phi_0^1}{\phi_0^1} + \braket{\phi_1^1}{\phi_1^1} + 2 Re(\braket{\phi_0^1}{\phi_1^1} ) \right],\\
p(\emptyset|+) &= \frac{1}{2} \left(\braket{\phi_0^{\emptyset}}{\phi_0^{\emptyset}} + \braket{\phi_1^{\emptyset}}{\phi_1^{\emptyset}} + 2 Re(\braket{\phi_0^{\emptyset}}{\phi_1^{\emptyset}}) \right).
\end{array}
\end{equation}

Now we estimate the phase error probability in the ideal case, where Alice is able to send the $\ket{+}$ and $\ket{-}$ states and Bob is able to measure both of them in the X basis.
In this case, following the formalism introduced before, the phase error rate can be expressed as a function of Eve's ancilla states as:  

\begin{multline}
e_x = \frac{p(-|+) + p(+|-)}{p(-|+) + p(+|-) + p(+|+) + p(-|-)} =\\
 1 - \frac{  2 Re(\braket{\phi_0^0}{\phi_1^1} + \braket{\phi_0^1}{\phi_1^0})}{\braket{\phi_0^0}{\phi_0^0} + \braket{\phi_1^0}{\phi_1^0} + \braket{\phi_0^1}{\phi_0^1} + \braket{\phi_1^1}{\phi_1^1}}.
\end{multline}

In our scenario where the POVM's elements on the monitoring line are defined following Eq.~(\ref{eq_POVM}), we can perfectly reproduce the given formula using the conditional probabilities of Eq.~(\ref{eq_pxz}) and Eq.~(\ref{eq_pxx}) (evaluated on measurable events). After a straightforward algebraic calculation it can be verified that the phase error rate in our protocol has the form given by Eq.~(\ref{eq_ex3}), where the second term is constrained to be positive since it corresponds to the probability $p(+|-)$.

%In order to use the number of detections corresponding to each event we have to pass from conditional probabilities to joint probabilities using Eq.~(\ref{eq_pjE}) and Eq.~(\ref{eq_pjL}) to obtain Eq.~(\ref{eq_ex4}).

\begin{figure}
\includegraphics[width = \columnwidth]{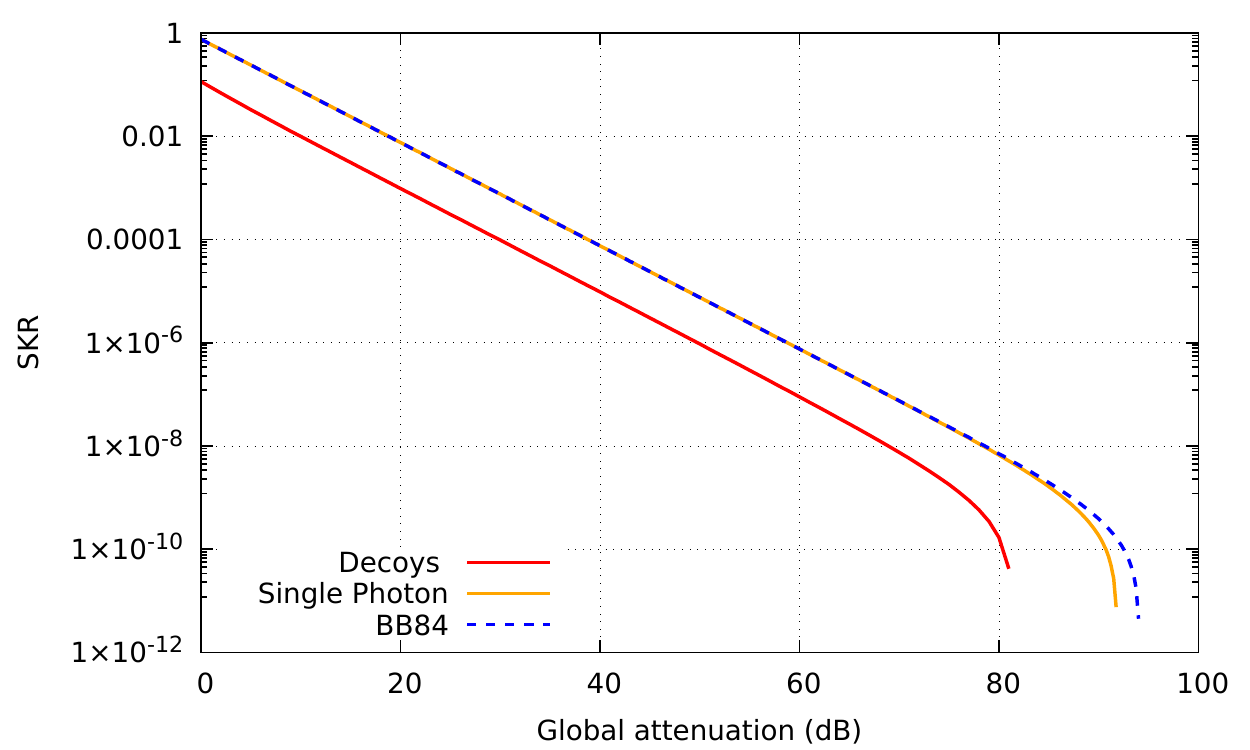}
\caption{\label{fig_SKR}Comparison between a standard BB84 protocol with a single-photon source (blue line) and our protocol with a single-photon source (yellow line) and phase-randomized weak coherent pulses (red line) with one decoy state \cite{rusca2018}. For all curves was considered an intrinsic error of 1\% and a probability of dark counts of $10^{-10}$  with security and correctness parameters ($\varepsilon_{sec}$ and $\varepsilon_{cor}$ respectively) equal to $10^{-9}$.}
\end{figure}

%We can now define the number of detections corresponding to each event as follows:

%\begin{equation}
%n(b,j,j) = N_{tot} p(b,j,j),
%\end{equation}
%
%where $N_{tot}$ is the total number of pulses sent by Alice.

Finally, in Fig.~\ref{fig_SKR} we show the achievable SKR of our protocol in comparison to a standard BB84 with a single-photon source. If we consider the single-photon source case, there is almost no difference except for really high attenuation. In this regime, where the errors increase due to the dark-counts of the detectors, our protocol is more affected, in the phase error estimation than a standard BB84 protocol. This is due to the fact that, in our protocol, there are three time-bins considered instead of two. However this difference is well compensated by the simplification of the implementation and allows, anyway, to achieve record breaking distances for QKD~\cite{boaron2018}. Moreover, by implementing our protocol with phase-randomized weak coherent pulses and one-decoy, we achieve a high SKR compared to that of single-photon sources (see Fig.~\ref{fig_SKR}).

\section{Conclusion and outlooks}

We have presented a simple and practical scheme that not only employs a limited amount of preparation states (three states and two pulse intensities when implemented with coherent pulses) but also allows us to use a simpler detection scheme. 
The next step in the security proof would be to introduce a complete analysis against coherent attacks. However we can already state that, since this protocol uses a phase-randomized source, techniques such as Azuma's inequality \cite{azuma1967,tamaki2009} and the quantum De Finetti's theorem \cite{konig2005} can be directly applied.

%%%%%%%%%%%%%%%%%%%%%%%%%%%%%%%%%%%%%%%%%%%%%%%%%%%%%%%%%%%%%%%%%%%%%%%%%%%%%%%%
\section*{Acknowledgements}
%%%%%%%%%%%%%%%%%%%%%%%%%%%%%%%%%%%%%%%%%%%%%%%%%%%%%%%%%%%%%%%%%%%%%%%%%%%%%%%%
We would like to acknowledge Charles Ci Wen Lim and Wailong Wang for the useful discussions about the security proof. We thank the Swiss NCCR QSIT, the EU's H2020 program under the Marie Sk\l{}odowska-Curie project QCALL (GA 675662), the Eurostars-2 joint programme (grant agreement: E11493 - QuPIC) and the Spanish Ministry of Economy and Competitiveness (MINECO), the Fondo Europeo de Desarrollo Regional (FEDER) through grants TEC2014-54898-R and TEC2017-88243-R.

\bibliography{Th_Stallion}
\end{document}